\documentclass{aastex61}

\received{July 1, 2016}
\revised{September 27, 2016}
\accepted{\today}
\submitjournal{ApJ}

\shorttitle{Outflows in  $z\sim0.4-0.8$ Quasars}
\shortauthors{Wang et al.}
\begin{document}

\title{A Study of Outflows in Luminous Quasars at Redshift $\sim0.4-0.8$}

\correspondingauthor{J. Wang}
\email{wj@nao.cas.cn}

\author[0000-0002-0786-7307]{J. Wang}
\affil{Key Laboratory of Space Astronomy and Technology, National Astronomical Observatories, Chinese Academy of Sciences, Beijing
100012, China}
\affil{School of Astronomy and Space Science, University of Chinese Academy of Sciences, Beijing, China}

\author{D. W. Xu		}
\affiliation{Key Laboratory of Space Astronomy and Technology, National Astronomical Observatories, Chinese Academy of Sciences, Beijing
100012, China}
\affiliation{School of Astronomy and Space Science, University of Chinese Academy of Sciences, Beijing, China}
%\collaboration{(AAS Journals Data Scientists collaboration)}

\author{J. Y. Wei}
\affiliation{Key Laboratory of Space Astronomy and Technology, National Astronomical Observatories, Chinese Academy of Sciences, Beijing
100012, China}
\affiliation{School of Astronomy and Space Science, University of Chinese Academy of Sciences, Beijing, China}
\nocollaboration



\begin{abstract}

We perform a systematic study of outflow in the narrow-line region (NLR) of active galactic nuclei (AGNs) at $z\sim0.4-0.8$  
basing upon a large sample of $\sim900$ quasars at $z\sim 0.4-0.8$. The sample is extracted from the Sloan Digital Sky Survey
by mainly requiring 1) the g-band magnitude is brighter than 19 magnitude; and 2) the [\ion{O}{3}]$\lambda5007$ 
emission line has a signal-to-noise ration larger than 30.
Profiles of multiple emission lines are modeled by a sum of several Gaussian functions.
The spectral analysis allows us to identify 1) a prevalence of both [\ion{O}{3}]$\lambda5007$ line blue asymmetry and 
bulk velocity blueshift of both [\ion{Ne}{3}]$\lambda3869$ and [\ion{Ne}{3}]$\lambda3426$ lines, 
when the [\ion{O}{2}]$\lambda3727$ line is used as a reference.
The velocity offset of [\ion{O}{3}]$\lambda5007$ line is, however, distributed around zero value, except for a few outliers.
2) not only the significant [\ion{O}{3}]$\lambda5007$ line asymmetry, but also the large bulk velocity offsets of [\ion{Ne}{3}]$\lambda3869$
and [\ion{Ne}{5}]$\lambda3426$ emission lines tend to occur in the objects with high $L/L_{\mathrm{Edd}}$,
which is considerably consistent with the conclusions based on local AGNs. With three $M_{\mathrm{BH}}$ estimation methods,
the significance level of the trend is found to be better than $2.9\sigma$, $3.2\sigma$ and $1.8\sigma$ for [\ion{O}{3}], [\ion{Ne}{3}] and [\ion{Ne}{5}],
respectively. \rm After excluding the role of radio jets, 
the revealed dependence of NLR gas outflow on $L/L_{\mathrm{Edd}}$ allows us to argue that  
the pressure caused by the wind/radiation launched/emitted from central supermassive black hole is the most likely origin of the 
outflow in these distant quasars, which implies that the outflow in luminous AGNs up to $z\sim1$ have the same origin.

\end{abstract}




\section{Introduction} \label{sec:intro}

A widely accepted idea on the growth of supermassive black holes (SMBHs) is the 
co-evolution of SMBHs and their host galaxies where the SMBHs reside in (see a 
recent review in Heckman \& Best 2014).
In the idea, a feedback process is generally necessary to self-regulate SMBH growth and star formation in the host
galaxy by either sweeping out circumnuclear gas in both galaxy merger and
secular evolution scenarios (e.g., Alexander \& Hickox 2012; Kormendy \& Ho 2013; Villar-Martin et al. 2016; Woo et al. 2017) 
or triggering star formation by compressing the gas (e.g., Zubovas et al. 2013; Ishibashi \& Fabian 2014).  
For instance, a suppressed star formation is revealed in high-z ($z=1\sim3$) powerful active galactic nuclei (AGNs) 
basing upon the Herschel SPIRE observations in sub-millimeter (Page et al. 2012). 
On the contrary, a positive feedback is revealed in the observations of individual objects
(e.g., Cresci et al. 2015; Carniani et al. 2016)

As suggested by both semianalytic models and
numerical simulations, the feedback process is required
not only to reproduce the observed $M_{\mathrm{BH}}-\sigma_\star$ relation,
luminosity functions of quasars, and normal galaxies (e.g.,
Haehnelt et al. 1998; Silk \& Rees 1998; Fabian 1999;
Kauffmann \& Haehnelt 2000; Granato et al. 2004; Di Matteo
et al. 2005, 2007; Springel et al. 2005; Croton et al. 2006;
Hopkins et al. 2007, 2008; Khalatyan et al. 2008; Menci et al.
2008; Somerville et al. 2008), but also to solve the ``over
cooling'' problem in the $\Lambda$ cold dark matter ($\Lambda$CDM) galaxy
formation model in which the cooling predicted in galaxy
groups and clusters is stronger than the observed one (e.g.,
Ciotti \& Ostriker 2007; Somerville et al. 2008, Hirschmann
et al. 2014). 
Dehnen \& King (2013)  proposed
a scenario of SMBH growth in which the SMBH accretion
disk is formed because the gas that is
swept-up by the feedback finally falls towards the SMBH on
near-parabolic orbit when the feedback weakens.

The occurrence of feedback from central SMBH can be diagnosed by the frequently observed 
outflows on various scales (see reviews in Veilleux et al. 2005
and Fabian 2012). The outflows can be traced by 
the blueshifted absorption lines in optical, UV, and soft-X-ray spectra (e.g., Crenshaw
et al. 2003; Hamann \& Sabra 2004; Wang \& Xu 2015 and
references therein). In addition, it can be more conveniently traced by
the blue asymmetry of strong [\ion{O}{3}]$\lambda\lambda$4959, 5007 doublet and its
bulk blueshift with respect to the local system (e.g., Heckman
et al. 1981; Veron-Cetty et al. 2001; Zamanov et al. 2002;
Marziani et al. 2003; Aoki et al. 2005; Bian et al. 2005;
Boroson 2005; Komossa et al. 2008; Xu \& Komossa 2009; Mullaney et al. 2013; Zhang et al. 2013; 
Harrison et al. 2014).

The origin of both blue asymmetry of the [\ion{O}{3}] doublet and its
bulk blueshift has been largely examined in past decades. 
However, discrepant conclusions have been drawn for AGNs at 
local universe.
Three kinds of modes, including AGN wind (e.g.,
Crenshaw et al. 2003; Pounds et al. 2003; Ganguly et al. 2007;
Reeves et al. 2009; Dunn et al. 2010; Tombesi et al. 2012),
radiation pressure (e.g., Granato et al. 2004; Alexander
et al. 2010), and mechanical energy outflow caused by a
collimated radio jet (e.g., Best et al. 2006; Holt et al. 2008;
Nesvadba et al. 2008; Rosario et al. 2010; Guillard et al. 2012), are proposed 
for a driving of the observed asymmetry and bulk blueshift.
On the one hand, ample studies indicate that strong blue asymmetry and large the bulk
velocity blueshift (especially the extreme ``blue outliers'' defined as 
the objects with [\ion{O}{3}] bulk blueshift larger than $250 \mathrm{km\ s^{-1}}$, Zamanov et al. 2002)
tend to occur in the AGNs with high Eddington ratio ($L/L_{\mathrm{Edd}}$, where 
$L_{\mathrm{Edd}}=1.26\times10^{38}(M_{\mathrm{BH}}/M_\odot)\ \mathrm{erg\ s^{-1}}$ is
the Eddington luminosity) (e.g., Bian et al. 2005; Boroson 2005;
Zhang et al. 2011; Wang et al. 2011, 2016; Wang 2015;  Zamanov
et al. 2002; Zhou et al. 2006; Komossa et al. 2008). 
A coevolution between the outflow traced by [\ion{O}{3}]$\lambda5007$ line and host galaxy has 
been revealed by Wang et al. (2011) and Wang (2015) for both local type I and II AGNs. The authors
argued that the coevolution is most likely driven by the coevolution of $L/L_{\mathrm{Edd}}$.  
On the other hand,  a dependence of the [\ion{O}{3}] line width on radio luminosity at 1.4 GHz 
($L_{1.4\mathrm{GHz}}$) has been reported for a
sample of flat-spectrum radio galaxies (e.g., Heckman
et al. 1984; Whittle 1985). This result was confirmed and reinforced by 
Mullaney et al. (2013) and Zakamska \& Greene (2014) for a large sample of both local 
type I and type II AGNs detected by the SDSS spectroscopic survey. 

Feedback from AGN is expected to be strong in high redshift
AGNs where the peaks of both AGN's activity and star formation
occur roughly coincident (e.g., Ishibashi et al., 2013). 
Extremely energetic outflows with a kinetic energy flux of 
$\sim 1-5\% L_{\mathrm{bol}}$ have been identified in four high-z luminous QSOs by using 
UV absorption lines as a tracer of outflow (e.g., Borguet et al. 2013; Arav et al. 2013 and references thererin).
With the spatially resolved maps of the kinematics of the ionized gas, galactic-scale outflows have
been identified in a number of radio-quiet and extremely radio-loud AGNs at $z\sim1-3$ (e.g., Alexander et al. 2010;
Nesvadba et al. 2006, 2008). The outflow at high redshift of $z= 6.4189$ was revealed by Maiolino et al. (2012) 
in SDSS\,J114816.64+525150.3 by a detection of the broad wing of the [\ion{C}{2}]$\lambda158\mu$m emission.

Even though the comprehensive studies have been carried out for the outflows of local AGNs, 
the outflows at early universe have been so far only focused on a number of 
objects with unusual properties. In this paper, we perform a systematic study on the outflow of $\sim900$ 
AGNs at intermediate redshift $z\sim0.4-0.8$, which is the largest sample at this redshift range so far, by focusing on their 
gas kinematics diagnosed by highly-ionized narrow emission lines.

The paper is organized as follows. The sample selection and
spectral analysis are presented in Sections 2 and 3, respectively.
The analysis and statistical results are shown in Section 4, and the
implications are discussed in Section 5. A $\Lambda$CDM cosmology
with parameters $H_0= 70\ \mathrm{km\ s^{-1} Mpc^{-1}}$, $\Omega_m=0.3$, and
$\Omega_{\Lambda}=0.7$ (Spergel et al. 2003) is adopted throughout the paper.

%By using the SDSS spectroscopic survey, Wang
%et al. (2011) and Wang (2015) recently suggested a coevolution
%between the feedback and host galaxy based on the fact that an
%AGN that has an [O III]OE?5007 emission line with stronger blue
%asymmetry tends to be associated with a younger stellar
%population.

\section{Sample Selection} \label{sec:style}

We start from the Sloan Digital Sky Survey Data Release 7 (SDSS DR7) catalog (Abazajian et al. 2009).
After focusing on the objects classified as QSOs (i.e., $spclass=3$) by the SDSS pipelines (Bromley et al.
1998; Glazebrook et al. 1998), we require the 
objects have 1) a redshift between 0.4 and 0.8. This requirement is used to ensure both [\ion{O}{3}]$\lambda$5007 
and \ion{Mg}{2}$\lambda2800$ emission lines can be covered by the SDSS spectroscopic wavelength range; 
2) a $g$-band magnitude brighter than 19 magnitude, which is
necessary for a proper modeling of the optical continuum. To model the emission line profile reliable, we further require 
the [\ion{O}{3}]$\lambda$5007 emission line has S/N$>$30. The uncertainty of the emission line $\sigma_l$ is estimated from 
$\sigma_l=\sigma_c N^{1/2}[1+\mathrm{EW}/(N\Delta)]^{1/2}$ (e.g., Perez-Montero \& Diaz 2003), where $\sigma_c$ is the standard 
deviation of the continuum in a box near the line, $N$ is the number of pixels used to integrate line flux, EW is the equivalent 
width of the line, and $\Delta$ is the wavelength dispersion in unit of $\mathrm{\AA\ pixel^{-1}}$.

With these selections, there are finally in total 981 entries in our sample.

\section{Spectral Modeling} \label{sec:floats}

The 1-Dimensional spectra of these AGNs are
analyzed by the IRAF\footnote{IRAF is distributed by National Optical Astronomy Observatory, which is operated
by the Association of Universities for Research in Astronomy, Inc., under cooperative
agreement with the National Science Foundation.} package. 
At the beginning, each spectrum is corrected for the Galactic extinction
basing upon the color excess $E(B-V)$ taken from the Schlegel, Finkbeiner, and Davies Galactic
reddening map (Schlegel et al. 1998). An $R_V=3.1$
extinction law (Cardelli et al., 1989) of the MilkyWay is adopted in the correction.
Each extinction-corrected spectrum is then shifted to the rest frame, along with the flux correction
due to the relativity effect, by using the redshift provided by the
SDSS pipelines.

\subsection{Continuum Modeling and Removal} \label{subsec:tables}

The continuum of each rest-frame spectrum is modeled by
a linear combination of three components: (1) a broken power law
from the central AGN, in which the wavelength of the break
point and the two spectral indices are not fixed in the
continuum modeling, (2) a template of both high-order Balmer
emission lines and a Balmer continuum from the broad-line
region (BLR), and (3) an empirical template of both optical and ultraviolet \ion{Fe}{2}
complex. The intrinsic extinction due to the host
galaxy is involved in the modeling by a   
galactic extinction curve with $R_V=3.1$. 
A $\chi^2$ minimization is iteratively performed over the whole SDSS spectroscopic wavelength range in 
the observe-frame, which roughly corresponds to a rest-frame wavelength range from 2700 to 6500\AA,
except for the regions with known emission lines (e.g., H$\beta$, H$\gamma$, H$\delta$, [\ion{O}{3}]$\lambda\lambda$4959, 5007,
[\ion{O}{2}]$\lambda$3727, [\ion{Ne}{3}]$\lambda3869$, [\ion{Ne}{5}]$\lambda3426$, and \ion{Mg}{2}$\lambda2800$).
The modeling and removal of the continuum is illustrated in the left panels of 
Figure 1 for two typical cases, one with weak optical \ion{Fe}{2} complex, and the another with strong \ion{Fe}{2} complex.
 
\begin{figure}[ht!]
\plotone{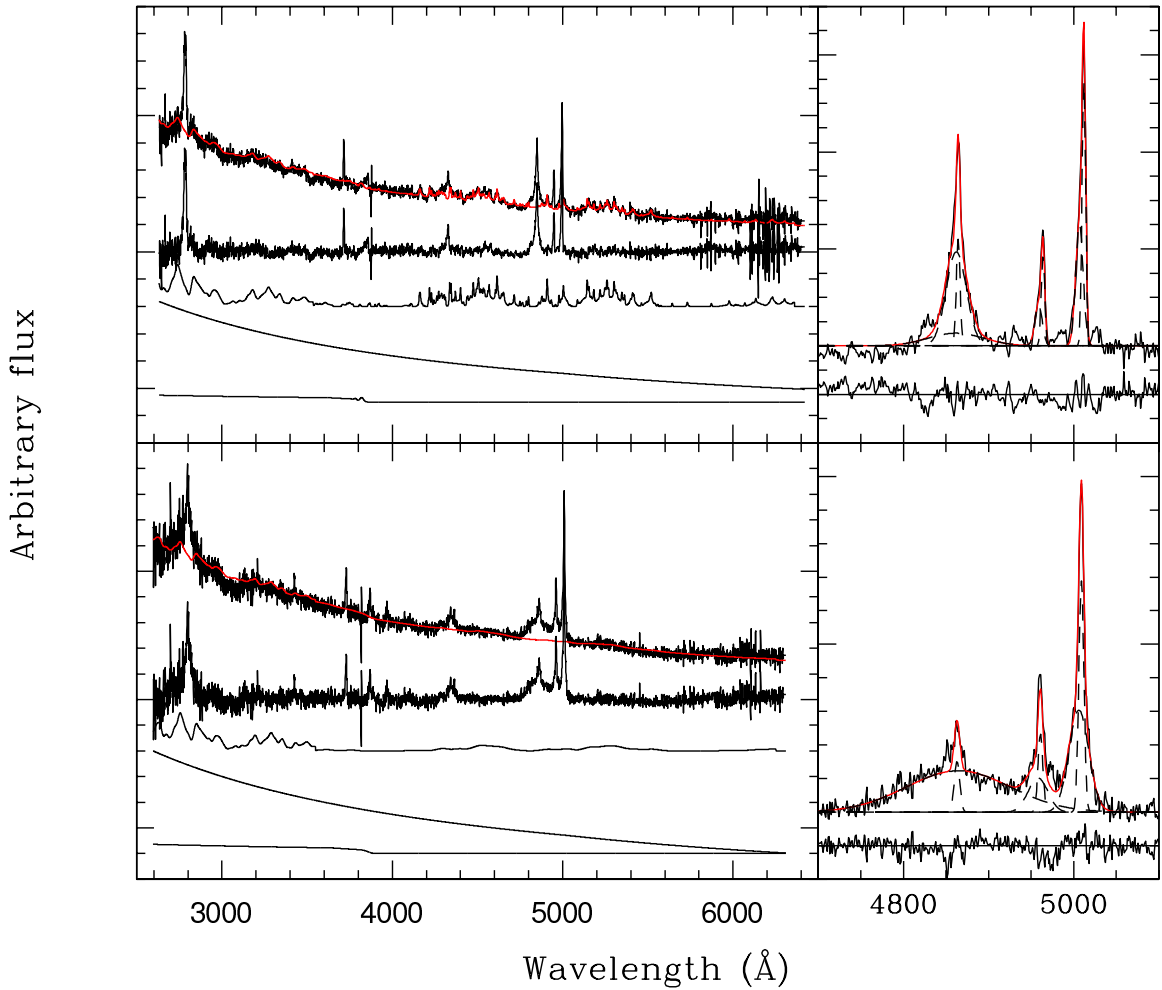}
\caption{\it Left panels: \rm Illustration of the modeling and removal of the continuum for two
typical cases. In each panel, the top curve shows the observed rest-frame
spectrum overplotted by the modeled continuum by the red curve. The
continuum-removed emission-line spectrum is shown below the observed one.
The modeled continuum is obtained by a reddened linear combination of 
the emission from both optical and UV \ion{Fe}{2} complex,
a broken power law from the central AGN, the Balmer continuum, and the
high-order Balmer emission lines, which are plotted in ordinals below the
emission-line spectrum. The intrinsic extinction is considered in the modeling
by using a galactic extinction curve with $R_V=3.1$. All of the spectra are
shifted vertically by an arbitrary amount for visibility. \it Right panels: \rm
Line profile modelings by a linear combination of a set of Gaussian functions for the H$\beta$
region, in which the modeled continuum has already been removed
from the original observed spectrum. The observed and modeled line profiles are plotted by black and red solid lines, respectively.
Each Gaussian function is shown by a dashed line. 
The sub-panel underneath the line spectrum presents the residuals between
the observed and modeled profiles.}
\end{figure}

The case B recombination model with an electron
temperature of $T_e=1.5\times10^4$K and an electron density of
$n_e = 10^{8-10}\ \mathrm{cm^{-3}}$ (Storey \& Hummer 1995) is adopted as 
the used template of the high-order Balmer lines (i.e., $\mathrm{H_7}$-$\mathrm{H_{50}}$) 
The line width of the high-order Balmer lines
is fixed to be that of the broad component of H$\beta$, which is
determined by our line profile modeling (see below). 
The Balmer continuum $f_\lambda^{\mathrm{BC}}$ is approximately modeled by the emission
from a partially optically thick cloud by following Dietrich
et al. (2002), see also in Grandi (1982) and Malkan \& Sargent
(1982):

\begin{equation}
  f_\lambda^{\mathrm{BC}}=f_\lambda^{\mathrm{BE}}B_\lambda(T_e)(1-e^{-\tau_\lambda}); \lambda\leq\lambda_{\mathrm{BE}}
\end{equation}
where $f_\lambda^{\mathrm{BE}}$ is the continuum flux at the Balmer edge
$\lambda_{\mathrm{BE}}=3464$\AA, and $B_\lambda(T_e)$ is the Planck function at an electron
temperature of $T_e=1.0\times10^4$K. $\tau_\lambda$ is the optical depth at
wavelength $\lambda$, which is related to the one at the
Balmer edge $\tau_{\mathrm{BE}}$ as $\tau_\lambda = \tau_{\mathrm{BE}}(\lambda/\lambda_{\mathrm{BE}})^3$.  
We adopt a typical value of $\tau_{\mathrm{BE}} = 0.5$ in the current
Balmer continuum fitting.

The optical and ultraviolet \ion{Fe}{2} complex in each rest-frame spectrum are modeled
by the empirical templates provided in Veron-Cetty et al. (2004) and by the 
theoretical template in Bruhweiler \& Verner (2008), respectively.   
For the optical template, both the broad
and narrow components of the \ion{Fe}{2} emission are included in
the modeling. The line widths of the broad and narrow
\ion{Fe}{2} emission are determined from the line profile modeling of
the H$\beta$ emission line (see below).
Bruhweiler \& Verner (2008) calculated a grid of ultraviolet \ion{Fe}{2}
emission spectra. The predicted spectrum giving the best fit to
the observed I\,ZW\,1 spectrum is adopted in the current study, which is calculated 
for $\log[n_{\mathrm{H}}/(\mathrm{cm^{-3}})] = 11.0$, $\log[\Phi_\mathrm{H}/(\mathrm{cm^{-2}\ s^{-1}})] = 20.5$, $\xi/(1\ \mathrm{km\ s
^{-1}}) = 20$, and 830
energy levels. Again, the template is broadened by convolution with
a Gaussian profile having the same width as the H$\beta$ broad
emission.

\subsection{Emission Line Profile Modeling}

\subsubsection{H$\beta$ and [\ion{O}{3}]$\lambda5007$}

After the removal of the continuum, we model the emission-line profiles 
of H$\beta$ and [\ion{O}{3}]$\lambda\lambda4959,5007$ of each spectrum by a linear 
combination of a set of several Gaussian profiles through the SPECFIT task
(Kriss 1994) in the IRAF packages. 
Generally speaking, a linear combination of two or three Gaussian profiles
are required to reproduce the observed [\ion{O}{3}]$\lambda\lambda$4959, 5007 line
profiles well in most of the cases. In addition, a linear combination of
two or three broad Gaussian functions is necessary for
adequately reproducing the observed broad Balmer line profile in
about a half of the sample. In each of these cases, 
a residual H$\beta$ line profile, which is obtained by
subtracting the modeled narrow-line component (including the
modeled forbidden lines) from the observed profile, is then
used to measure the line width and integrated line flux of the
broad H$\beta$ emission. 
In the modelings, the line flux ratio of the [\ion{O}{3}]
doublet is fixed to its theoretical value. 
The line width of the narrow H$\beta$ component is fixed to equal that
of the [\ion{O}{3}] core line, if the resulted two widths are
significantly different. In order to properly model the
[\ion{O}{3}]$\lambda$5007 line profile, a broad \ion{He}{1}$\lambda$5016 emission line
(Veron et al. 2002) is additionally required in the line profile
modelings in a few objects.
The line modelings are schematically presented in 
the right panels of Figure 1 for the H$\beta$ and [\ion{O}{3}]$\lambda\lambda$4959, 5007 emission lines.

\subsubsection{Other Emission Lines}

With the similar technique, a combination of a single Gaussian profile and 
a local linear pseudo-continuum is adopted to model [\ion{O}{2}]$\lambda$3727, [\ion{Ne}{3}]$\lambda3869$ 
and [\ion{Ne}{5}]$\lambda3426$ forbidden lines through a $\chi^2$ minimization. The level of the 
pseudo-continuum is determined by both red and blue emission just outside individual emission line, which accounts for 
the residual resulted from our continuum removal aforementioned in Section 3.1, because the three forbidden lines
are typically relatively weak in the spectra.

\vspace{1cm}

In summary, our spectral analysis of the H$\beta$ region is available for 884 quasars at redshift between 0.4 and 0.8, 
except for three objects whose H$\beta$ emission is too weak to be modeled. 
There are in total 782 objects whose [\ion{O}{2}]$\lambda3727$ line emission has S/N$>$5. With the same criterion, the total entries of 
available measurements of [\ion{Ne}{3}]$\lambda$3869 and [\ion{Ne}{5}]$\lambda$3426 are 570 and 329, respectively.

\section{Analysis and Results}

\subsection{[\ion{O}{3}]$\lambda$5007 Line Asymmetry}

By modeling the [\ion{O}{3}]$\lambda$5007 emission-line profile by a sum of
several Gaussian profiles, we follow Wang et al. (2016) to parametrize the 
asymmetry of the [\ion{O}{3}] line by a velocity of $\delta\upsilon$, defined as
\begin{equation}
 \delta\upsilon=\frac{\Sigma_{k=1}^n f_k(\upsilon_k-\upsilon_p)}{\Sigma_{k=1}^n f_k}
\end{equation}
where $f_k$ and $\upsilon_k$ is the modeled flux and velocity of the kth
Gaussian function, respectively. $\upsilon_p$ denotes the velocity of the
Gaussian profile that reproduces the peak of the observed line
profile. A negative value of $\delta\upsilon$ corresponds to a blue asymmetry, and a
positive one corresponds to a red asymmetry. Wang et al. (2016) has shown that 
the asymmetry parameter defined in Equation (2) is related with the parameter defined 
in Harrison et al. (2014) quiet well, which quantifies the line
asymmetry by comparing the measured line widths/centers (in units of either
wavelength or velocity) at different line flux levels (e.g, Heckman et al.
1981; Whittle 1985; Veilleux 1991; Wang et al. 2011; Liu et al.
2013), even through there is a small systematical difference
between the two definitions.

\subsection{Bulk Velocity Shifts of Forbidden Lines}

We calculate bulk relative velocity shifts of high-ionized emission lines (i.e., [\ion{O}{3}]$\lambda5007$, 
[\ion{Ne}{3}]$\lambda3869$, and [\ion{N}{5}]$\lambda3426$) with respect to the [\ion{O}{2}]$\lambda3727$ emission line.
The [\ion{O}{2}] emission line is adopted as a velocity shift reference in the current study for two reasons. 
At first, although the narrow H$\beta$ emission line is widely used as a reference in a great deal of previous studies,
recent studies indicate that the narrow Balmer lines show an evident velocity shift relative to
the galaxy rest frame. Hu et al. (2008) suggested
that the [\ion{O}{2}] emission line might be a more reliable reference than
either [\ion{O}{3}] or H$\beta$ in local type I AGNs. 
In a large sample of $\sim$23,000 local type II AGNs, Bae \& Woo (2014) found that 
H$\alpha$ shows large velocity offsets being comparable to that of [\ion{O}{3}] in $\sim3\%$ of the sample when the 
host galaxy is adopted as a reference. A large velocity shift of $\sim200\ \mathrm{km\ s^{-1}}$ in the host-galaxy frame is identified for 
both narrow H$\alpha$ and H$\beta$ lines in SDSS\,J112611.63+425246.4, a Balmer-absorption AGN, by Wang \& Xu (2015).
Secondly, the poor H$\beta$ profile prohibits an available extraction of the H$\beta$ narrow component in $\sim20\%$ objects listed
in our sample. 

The finally used bulk relative velocity shift is defined
as $\Delta\upsilon=c\Delta\lambda/\lambda_0$, where $\lambda_0$ and $\Delta\lambda$ are
the rest-frame wavelength in vacuum of a given emission line and 
the wavelength shift of the line with respect to the [\ion{O}{2}]$\lambda3727$ emission line, respectively. 
$\Delta\lambda$ is obtained from the modeled line centers as $\Delta\lambda=(\lambda^{\mathrm{ob}}-\lambda^{\mathrm{ob}}_{\mathrm{[OII]}})-
(\lambda_0-\lambda_{0,\mathrm{[OII]}})$, where $\lambda^{\mathrm{ob}}$ ($\lambda^{\mathrm{ob}}_{\mathrm{[OII]}}$)
and $\lambda_0$ ($\lambda_{0,\mathrm{[OII]}}$) are the observed line center resulted from our profile modeling and the
line wavelength in vacuum of a given ([\ion{O}{2}]) line, respectively.
A negative value of $\Delta\upsilon$ corresponds to a blueshift, and a positive value
corresponds to a redshift.

%% Note that the \setcounter and \renewcommand are needed here because
%% this example is using a mix of deluxetable and tabular.  Here the
%% deluxetable counters are set with \tablenum but the situation is a bit
%% more complex for tabular.  Use the first command to set the Table number
%% to ONE LESS than it should be.  The next command will auto increment it
%% to the desired number.
\subsection{Velocity Dispersion of Broad H$\beta$ Emission Line}

We measure the velocity dispersion of the broad H$\beta$ component $\sigma_{\mathrm{H\beta}}$ of each object from the 
H$\beta$ broad line profile obtained by subtracting the modeled narrow-line component. The dispersion is based on 
the definition given in Peterson et al. (2004):
\begin{equation}
  \sigma_{\mathrm{H\beta}}^2=\frac{\int\lambda^2f_\lambda d\lambda}{\int f_\lambda d\lambda}-\overline{\lambda}^2
\end{equation}
where $f_\lambda$ is the line specific flux after a subtraction of the continuum and 
$\overline{\lambda}=\int\lambda f_\lambda d\lambda/\int f_\lambda d\lambda$ is the 
first moment of the line profile.

\subsection{Estimation of $M_{\mathrm{BH}}$ and $L/L_{\mathrm{Edd}}$} \label{subsubsec:autonumber}

Both SMBH mass ($M_{\mathrm{BH}}$) and $L/L_{\mathrm{Edd}}$ are the critical parameters
describing AGN phenomena (e.g., Shen \& Ho 2014). 
The great progress made in the
reverberation mapping technique (e.g., Kaspi et al. 2000, 2005; 
Peterson \& Bentz 2006; and see Marziani \& Sulentic 2012 and
Peterson 2014 for recent reviews; Wang et al. 2014; Du et al. 2015) allows us to 
estimate $M_{\mathrm{BH}}$ (and also $L/L_{\mathrm{Edd}}$) basing upon the broad emission line in a single 
epoch spectroscopy (e.g., Wu et al. 2004).

$M_{\mathrm{BH}}$ is estimated from the modeled broad H$\beta$ line emission 
for all the 884 intermediate-z quasars, except for the three objects without available H$\beta$ profile modeling, 
according to several calibrated
relationships.

At first, the estimation of $M_{\mathrm{BH}}$ provided in Vestergaard \& Peterson (2006, and references
therein) is given as 
\begin{equation}
 M_{\mathrm{BH}}=10^{6.67}\bigg(\frac{L_{\mathrm{H\beta}}}{10^{42}\ \mathrm{erg\ s^{-1}}}\bigg)^{0.63}\bigg(\frac{\mathrm{FWHM(H\beta})}{1000\ \mathrm{km\ s^{-1}}}\bigg)^2\ M_\odot
\end{equation}
where $L_{\mathrm{H\beta}}$ is the luminosity of the H$\beta$ broad
component and FWHM(H$\beta$) is the line width of broad H$\beta$ emission that is resulted from our line
profile modeling (Section 3.2.1). 
Secondly, an alternative $M_{\mathrm{BH}}$ estimation based on the line dispersion $\sigma_{\mathrm{H\beta}}$ 
is obtained through the method presented in Collin et al. (2006) who argued that 
the $M_{\mathrm{BH}}$ estimated from line dispersion is less biased than that from FWHM.
Finally, we obtain an additional $M_{\mathrm{BH}}$ estimation in which the impact of radiation pressure on the 
kinematics of BLR gas is taken into account of by following the calibration reported in Marconi et al. (2008)
\begin{equation}
\tiny
  \frac{M_{\mathrm{BH}}}{M_\odot}=\tilde{f}\bigg(\frac{\mathrm{FWHM(H\beta)}}{1000\ \mathrm{km\ s^{-1}}}\bigg)^2\bigg(\frac{L_{\mathrm{5100}}}{10^{44}\ \mathrm{erg\ s^{-1}}}\bigg)^{0.5}
+g\bigg(\frac{L_{\mathrm{5100}}}{10^{44}\ \mathrm{erg\ s^{-1}}}\bigg)
\end{equation}
where the logarithmic of the two parameters $\tilde{f}$ and $g$ are $6.13^{+0.15}_{-0.30}$ and $7.72^{+0.06}_{-0.05}$, respectively.
 $L_{5100}$ is the AGN's continuum luminosity at 5100\AA\ that is inferred from H$\beta$ line luminosity 
through the calibration given in Greene \& Ho (2005)  
\begin{equation}
  L_{5100}=7.31\times10^{43}\bigg(\frac{L_{\mathrm{H\beta}}}{10^{42}\ \mathrm{erg\ s^{-1}}}\bigg)^{0.883}\ \mathrm{erg\ s^{-1}}
\end{equation}

In order to obtain $L_{\mathrm{bol}}/L_{\mathrm{Edd}}$, the bolometric luminosity $L_{\mathrm{bol}}$ 
of each object is estimated from the usually used bolometric correction $L_{\mathrm{bol}}=9\lambda L_\lambda(5100\mathrm{\AA})$ 
(Kaspi et al. 2000). \rm
Note that, in the above estimations the intrinsic extinction is not applied for $L_{\mathrm{H\beta}}$ because 
the Balmer decrement that is traditionally used is either unavailable or unreliable for most of the objects.

\subsection{Statistical Results} \label{subsubsec:hide}

With the quantification of line asymmetry, bulk velocity offset, $M_{\mathrm{BH}}$ and $L/L_{\mathrm{Edd}}$, the main statistical results 
are presented in this subsection.

\subsubsection{Outflow and SMBH Accretion}

\it FWHM(H$\beta$) versus outflow.\rm \ 
The top line in Figure 2 plots the line width of H$\beta$ broad emission as 
a function of [\ion{O}{3}] line asymmetry $\delta\upsilon$,  bulk velocity offsets $\Delta\upsilon$
of [\ion{O}{3}], [\ion{Ne}{3}] and [\ion{Ne}{5}] from left to right. 
%We note at first that our spectral analysis indicates a prevalence of [\ion{O}{3}]$\lambda5007$ line blue asymmetry, and 
%bulk velocity blueshift of both [\ion{Ne}{3}]$\lambda3869$ and [\ion{Ne}{3}]$\lambda3426$ lines. It is interesting that 
%the velocity offset of [\ion{O}{3}]$\lambda5007$ line is distributed around zero value, except for a few outliers with 
%narrow H$\beta$ broad components.   
The left panel in the top line shows that 1)
blue asymmetry is quite prevalent for the [\ion{O}{3}] emission line, which is consistent with previous results concluded from 
the samples of local type I and II AGN's (e.g., Harrison et al. 2014; Mullaney et al. 2013; Wang et al. 2011, 2016); and
2) stronger the blue asymmetry, smaller the $\mathrm{FWHM(H\beta)}$ (i.e., $\mathrm{FWHM(H\beta)}<5000\ \mathrm{km\ s^{-1}}$ ) will be. 
One can see from the $\mathrm{FWHM(H\beta)}$ versus $\Delta\upsilon_{\mathrm{[OIII]}}$ plot that the bulk velocity offset of [\ion{O}{3}]
line is almost asymmetrically distributed around zero value\footnote{In fact, when the host galaxy is used as a reference, 
Wang \& Xu (2016) revealed a marginal bulk blueshift of [\ion{O}{3}] emission line of $-10\pm30\ \mathrm{km\ s^{-1}}$ in a case study of 
Balmer-absorption AGN SDSS\,J112611.63+425246.4}, except for a few outliers with extremely large bulk blueshift.
The outliers typically have small $\mathrm{FWHM(H\beta)}<5000\ \mathrm{km\ s^{-1}}$, which agrees with previous studies on local AGNs in quality.
The local extreme ``blue outliers'', usually defined as the objects with
[\ion{O}{3}] bulk blueshift larger than $250\ \mathrm{km\ s^{-1}}$, are found
to dominantly occur in the AGNs with small $\mathrm{FWHM(H\beta)}$ (e.g., Zamanov
et al. 2002; Zhou et al. 2006; Komossa et al. 2008). Not as similar as the case of $\Delta\upsilon[\mathrm{OIII}]$, both 
[\ion{Ne}{3}] and [\ion{Ne}{5}] emission lines show a dominance of bulk velocity blueshift whose value decreases 
with $\mathrm{FWHM(H\beta)}$.  

\it H$\beta$ Line profile versus outflow. \rm 
The panels in the second line of Figure 2 is the same as the top one, but for $\mathrm{FWHM/\sigma_{H\beta}}$ that depends on the line profile 
caused by a variety of line-of-sight kinematics and/or inclination of BLR gas. We theoretically have 
$\mathrm{FWHM/\sigma_{H\beta}}=2\sqrt{2\ln2}=2.35$ for a pure Gaussian profile, and $\to 0$ for a pure Lorentzian profile. The 
four panels in the line show that the measured outflow velocities are independent on the $\mathrm{FWHM/\sigma_{H\beta}}$, i.e., kinematics and/or inclination of BLR,
although there seems to be a weak tendency for the case of [\ion{Ne}{5}] in which large velocity offset tends to occur in the objects
with stronger line wings (i.e., $\mathrm{FWHM/\sigma_{H\beta}}\sim1.5$).

\it $L/L_{\mathrm{Edd}}$  versus outflow. \rm
The $L/L_{\mathrm{Edd}}$ based on the three $M_{\mathrm{BH}}$ estimations are plotted against the measured outflow velocities in 
the panels from lines 3 to 5 in Figure 2. A comparison of the three estimations of $L/L_{\mathrm{Edd}}$ allows us to draw 
a robust conclusion: \rm
Aside from the bulk velocity offset of [\ion{O}{3}], 
objects with stronger [\ion{O}{3}] line blue asymmetry, larger bulk blueshift of both [\ion{Ne}{3}] and [\ion{Ne}{5}] emission lines 
tend to be associated with higher $L/L_{\mathrm{Edd}}$, although an object with large $L/L_{\mathrm{Edd}}$ does not necessarily
have strong [\ion{O}{3}] line blue asymmetry and large bulk blueshift possibly due to the orientation of the outflow.   
Finally, we emphasize that no dependence can be found when $L/L_{\mathrm{Edd}}$ is taken place by 
either H$\beta$ (and [\ion{O}{3}]) luminosity or $M_{\mathrm{BH}}$.

\vspace{0.5cm}
Table 1 lists the corresponding correlation coefficient matrix that is obtained from the Spearman rank-order test. For each 
correlation coefficient, the corresponding probability of the null correlation is shown in the
bracket. We emphasize that the weak correlation significance shown in the table has no essential impact on the 
conclusion of this study. Our conclusion states that a high $L/L_{\mathrm{Edd}}$ is a necessary, rather than a sufficient, condition for 
significant outflow, because the measured outflow velocity additionally depends on the orientation of the outflow with respect to
the line-of-sight of an observer.  

%\begin{figure}[ht!]
\begin{figure*}
\plotone{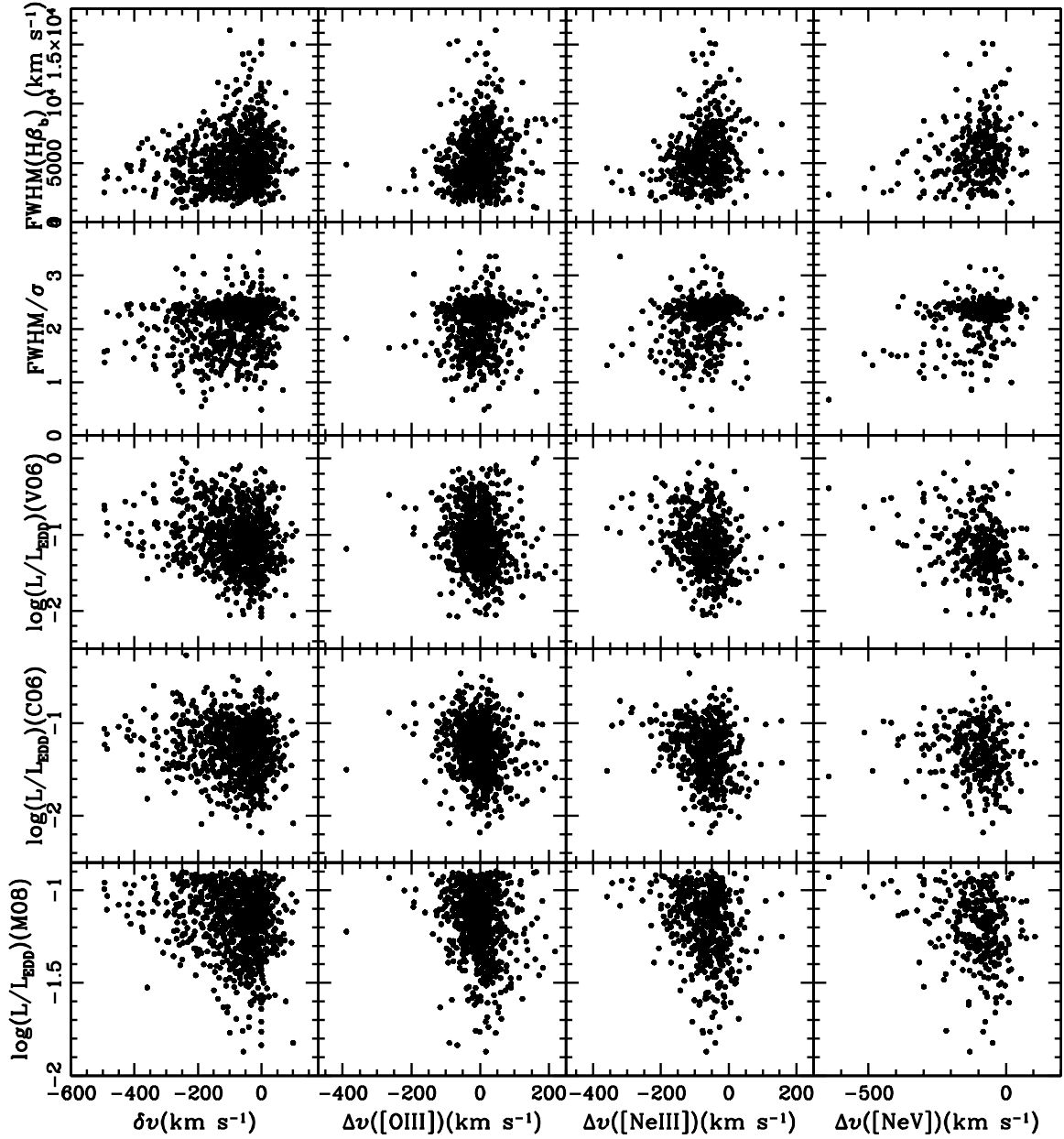}
\caption{\it Line 1: \rm from left to right, line width of H$\beta$ broad emission (FWHM) is plotted as a function of [\ion{O}{3}]$\lambda5007$ line asymmetry 
$\delta\upsilon$ defined in Equation (2), bulk velocity shift $\Delta\upsilon$ of [\ion{O}{3}]$\lambda5007$, [\ion{Ne}{3}]$\lambda3869$ and [\ion{Ne}{5}]$\lambda3426$
emission lines. All the  $\Delta\upsilon$ are determined by using [\ion{O}{2}]$\lambda3727$ line as a reference. 
A positive velocity denotes a redshift, and a negative value a blueshift. \it Line 2: the same as line 1, but for $\mathrm{FWHM/\sigma_{\mathrm{H\beta}}}$, where 
$\sigma_{\mathrm{H\beta}}$ is estimated based on the definition in Eq. (3). \it Lines 3-5: the same as line 1, but for the $L/L_{\mathrm{Edd}}$ estimated through 
three calibrations, see Section 4.4 for the details.  
}
\end{figure*}

\begin{table*}
\begin{center}
\caption{Correlations coefficient matrix related with outflow velocities measured from the forbidden emission lines.\label{tbl-1}}
\begin{tabular}{lcccc}
\tableline\tableline
Property  & $\delta\upsilon$([\ion{O}{3}]) & $\Delta\upsilon$([\ion{O}{3}]) & $\Delta\upsilon$([\ion{Ne}{3}]) & $\Delta\upsilon$([\ion{Ne}{5}])\\
 & (1) & (2) & (3) & (4) \\ 
\tableline
(1) $\mathrm{FWHM(H\beta)}$   &  0.184($<10^{-4}$) &  0.179($<10^{-4}$) &  0.223($<10^{-4}$) & 0.218($<10^{-4}$)\\
(2) $L/L_{\mathrm{Edd}}$(V06) & -0.183($<10^{-4}$) & -0.198($<10^{-4}$) & -0.249($<10^{-4}$) & -0.288($<10^{-4}$)\\
(3) $L/L_{\mathrm{Edd}}$(C06) & -0.122($<10^{-4}$) & -0.177($<10^{-4}$) & -0.205($<10^{-4}$) & -0.165($<10^{-4}$)\\
(4) $L/L_{\mathrm{Edd}}$(M08) & -0.173($<10^{-4}$) & -0.203($<10^{-4}$) & -0.261($<10^{-4}$) & -0.233($<10^{-4}$)\\      
\tableline
%\tablenotetext{a}{The statistics is based on the 20 sources with detected radio flux taken from the FIRST catalog.}
%\tablenotetext{b}{The statistics is based on all the 47 sources through survival analysis. }
\end{tabular}
\end{center}
\end{table*}

\begin{figure*}
\plotone{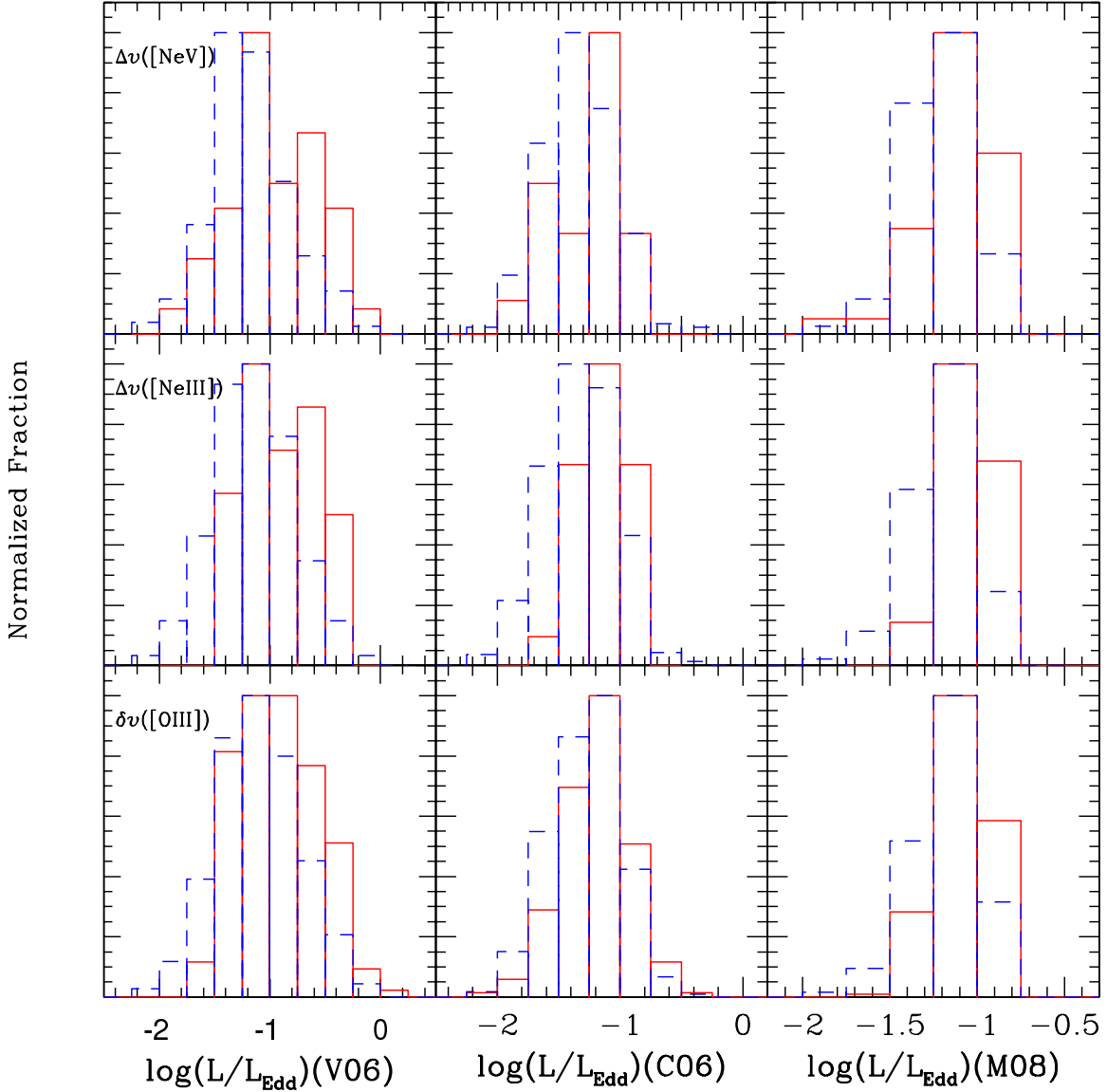}
\caption{ Comparisons of the normalized distributions of the $L/L_{\mathrm{Edd}}$ between the high-velocity and low-velocity groups.
The panels at the bottom are for $\delta\upsilon$([\ion{O}{3}]) for the $L/L_{\mathrm{Edd}}$ estimated from the three different methods (see Section 4.4
for the details.).  The middle and top panels are the same as the bottom ones, but for $\Delta\upsilon$(\ion{Ne}{3}) and $\Delta\upsilon$(\ion{Ne}{5}),
respectively. In each panel, the red-solid and blue-dashed lines denote high-velocity and low-velocity groups, respectively. The two groups are 
separated at a velocity of $-150\ \mathrm{km\ s^{-1}}$ for  $\delta\upsilon$([\ion{O}{3}]) and $\Delta\upsilon$(\ion{Ne}{3}), and at 
a velocity of $-200\ \mathrm{km\ s^{-1}}$ for $\Delta\upsilon$(\ion{Ne}{5}).  
}
\end{figure*}

As an additional test, we separate the entire sample into two parts basing upon the measured line profile asymmetry and 
bulk velocity shifts, i.e., $\delta\upsilon(\mathrm{[OIII]})=-150\mathrm{km\ s^{-1}}$, $\Delta\upsilon(\mathrm{[NeIII]})=-150\mathrm{km\ s^{-1}}$
and $\Delta\upsilon(\mathrm{[NeIII]})=-200\mathrm{km\ s^{-1}}$.
Figure 3 compares the normalized distributions of $L/L_{\mathrm{Edd}}$ estimated from the three methods between the two parts.   
For each of the panels, a two-sided Kolmogorov-Smirnov test is performed to check whether the high-velocity and low-velocity parts
are drawn from the sample population. The results of the Kolmogorov-Smirnov tests are listed in Table 2. Each entry contains 
the probability that the two samples are drawn from the same parent population, along with the maximum absolute discrepancy.

The statistical tests indicate that, for the $\delta\upsilon(\mathrm{[OIII]})$ case, 
we can reject the hypothesis that the two $L/L_{\mathrm{Edd}}$ distributions are drawn from the same parent population at a significance 
level better than $3\sigma$ (assuming a Gaussian function) when the $M_{\mathrm{BH}}$ is estimated from Vestergaard \& Peterson (2006) and 
Marconi et al. (2008), although the significance level deceases to $2.9\sigma$ when the method in Collin et al. (2006) is adopted. 
A similar trend could be found in the $\Delta\upsilon(\mathrm{[NeV]})$ case, in which 
the method in Collin et al. (2006) results in the lowest significance level of $1.8\sigma$. On the contrary, in the  $\Delta\upsilon(\mathrm{[NeIII]})$ case, 
the same hypothesis can be rejected at a significance 
level better than $3\sigma$ for all the three $M_{\mathrm{BH}}$ estimation methods, in which the best result ($3.7\sigma$) is obtained 
when the method in Collin et al. (2006) is adopted.

\begin{table*}
\begin{center}
\caption{Two-sided Kolmogorov-Smirnov Test Matrix.\label{tbl-2}}
\begin{tabular}{lccc}
\tableline\tableline
Property  & $\delta\upsilon$([\ion{O}{3}])  & $\Delta\upsilon$([\ion{Ne}{3}]) & $\Delta\upsilon$([\ion{Ne}{5}])\\
 & (1) & (2) & (3)  \\ 
\tableline
(1) $L/L_{\mathrm{Edd}}$(V06) & $8.1\times10^{-5}$(0.19) & $1.3\times10^{-3}(0.28)$ & $4.7\times10^{-3}$(0.29) \\
(2) $L/L_{\mathrm{Edd}}$(C06) & $4.1\times10^{-3}$(0.15) & $2.0\times10^{-4}(0.32)$ & $6.9\times10^{-2}$(0.22) \\
(3) $L/L_{\mathrm{Edd}}$(M08) & $2.2\times10^{-5}$(0.20) & $8.5\times10^{-4}(0.29)$ & $2.9\times10^{-3}$(0.30) \\      
\tableline
%\tablenotetext{a}{The statistics is based on the 20 sources with detected radio flux taken from the FIRST catalog.}
%\tablenotetext{b}{The statistics is based on all the 47 sources through survival analysis. }
\end{tabular}
\end{center}
\end{table*}

\subsubsection{$L/L_{\mathrm{Edd}}$ vs. $\delta\upsilon$: [OIII] Luminosity on kpc Scale or Outflow}
We obtained a dependence of $L/L_{\mathrm{Edd}}$ on the [\ion{O}{3}] line asymmetry measured by the parameter 
$\delta\upsilon$ defined in Eq. (2) in Section 4.5.1. One would argue that a large asymmetry is caused by a faint [\ion{O}{3}] 
emission from the narrow emission-line region (NLR) on kpc scale with a kinematics under the gravitation of the host galaxy 
rather than a strong emission from outflow gas. Additional test is therefore performed to test the potential impact of 
this effect as follows. At the beginning, since most of the objects show a blue [\ion{O}{3}] asymmetry, we obtain the 
[\ion{O}{3}] luminosity of NLR in kpc scale $L_{\mathrm{[OIII],NLR}}$ of each object through a scale\footnote{We do not use the
the modeled narrow core component given by the spectral profile modeling described in Section 3.2.1, because
based on our experience the modeled core
component is quite unreliable in the case with a smooth, significant blue line wing, even though the total profile
and flux can be modeled quite well with multiple Gaussian components.} 
$L_{\mathrm{[OIII],NLR}}=2L_{\mathrm{[OIII],red}}$, where $L_{\mathrm{[OIII],red}}$ is the line luminosity redward of the 
line peak, although we admit that this procedure potentially yields an over-estimation of the $L_{\mathrm{[OIII],NLR}}$.  
The left panel in Figure 4 presents the distribution of $L_{\mathrm{[OIII],NLR}}$.

With the distribution, the total sample is then separated into three groups according to their $L_{\mathrm{[OIII],NLR}}$: 
Group 1 has $\log L_{\mathrm{[OIII],NLR}}<42.5$, Group 2 has $42.5\leq\log L_{\mathrm{[OIII],NLR}}<43.0$ and 
Group 3 has $\log L_{\mathrm{[OIII],NLR}}\geq43.0$. All the three groups are over-plotted in the 
$L/L_{\mathrm{Edd}}$ vs. $\delta\upsilon$ diagram in the right panel of Figure 4. A comparison of the distributions
on the diagram suggests a consistence for all the three groups, which indicates that our conclusion on the 
$L/L_{\mathrm{Edd}}$ vs. $\delta\upsilon$ dependence is indeed resulted from a change of outflow luminosity rather than a change of 
NLR luminosity in kpc scale. A two-dimensional Kolmogorov-Smirnov test (Peacock 1983; Fasano \& Franceschini 1987)
yields a significance level of $<4.5\times10^{-3}$ for the hypothesis that the three groups are from different populations.

\begin{figure}
\plotone{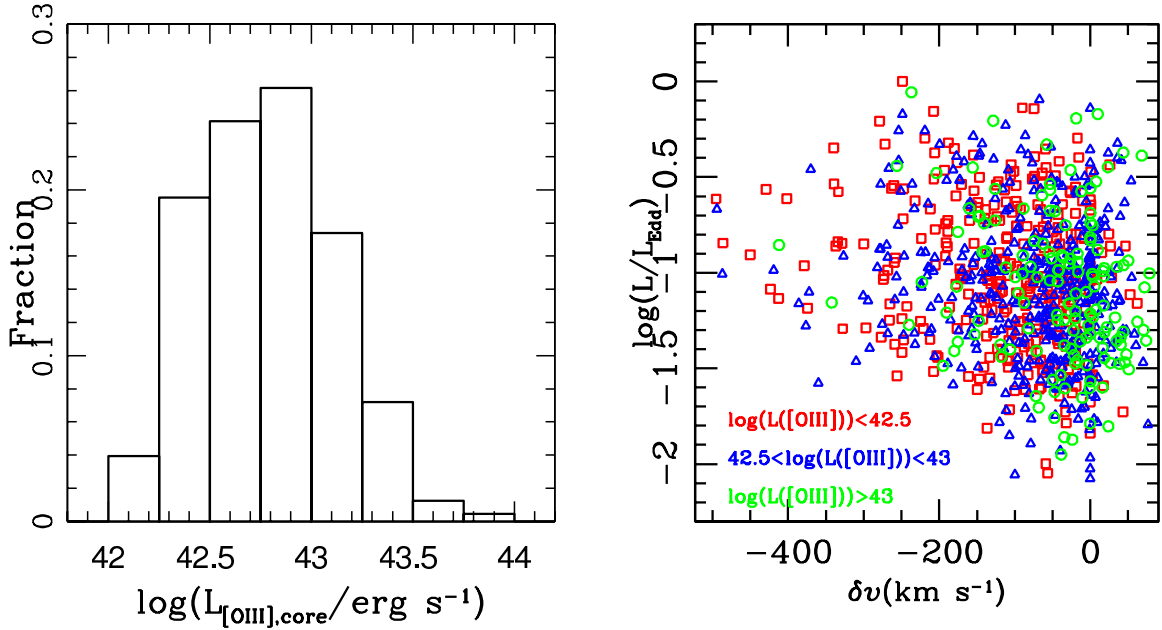}
\caption{\it Left panel: \rm Distribution of the [\ion{O}{3}]$\lambda5007$ line luminosity from NLR at kpc scale, which is obtained in 
Section 4.5.2. \it Right panel: \rm $L/L_{\mathrm{Edd}}$ versus $\delta\upsilon$ diagram, in which the total sample is separated into three 
groups according the their NLR [\ion{O}{3}] line luminosities. The three groups are over-plotted in the same diagram by different colors and symbols.    
}
\end{figure}

%$L'_{\mathrm{[OIII],out}}=L_{\mathrm{[OIII]}}-L_{\mathrm{[OIII],NLR}}$, where $L_{\mathrm{[OIII]}}$ and $L_{\mathrm{[OIII],NLR}}$
%are the total line luminosity obtained from our line profile modeling in Section      

\subsubsection{NLR Stratification}

The AGN's NLR gas has been generally believed to be
stratified in density and ionization potential for a long time (e.g., Filippenko \& Halpern 1984; Filippenko 1985; 
De Robertis \& Osterbrock 1986). 
A systematic stratification of kinematics of NLR gas is illustrated in Figure 5 for the studied $z\sim0.4-0.8$ quasars, 
which compares the bulk velocity offset with 
the ionization potential (IP) for three different emission lines: more highly ionized emission line tends to be associated with larger 
bulk blueshift, which is consistent with previous studies (e.g., Komossa et al. 2008; Wang \& Xu 2016). The red squares and 
errorbars in the plot represents the average value and the corresponding standard deviation at a confidence level of $1\sigma$,   
which yields a best-fit relation of $\Delta\upsilon=(106.6\pm152.9)-(2.3\pm2.3)\mathrm{IP}$.

\begin{figure}[ht!]
\plotone{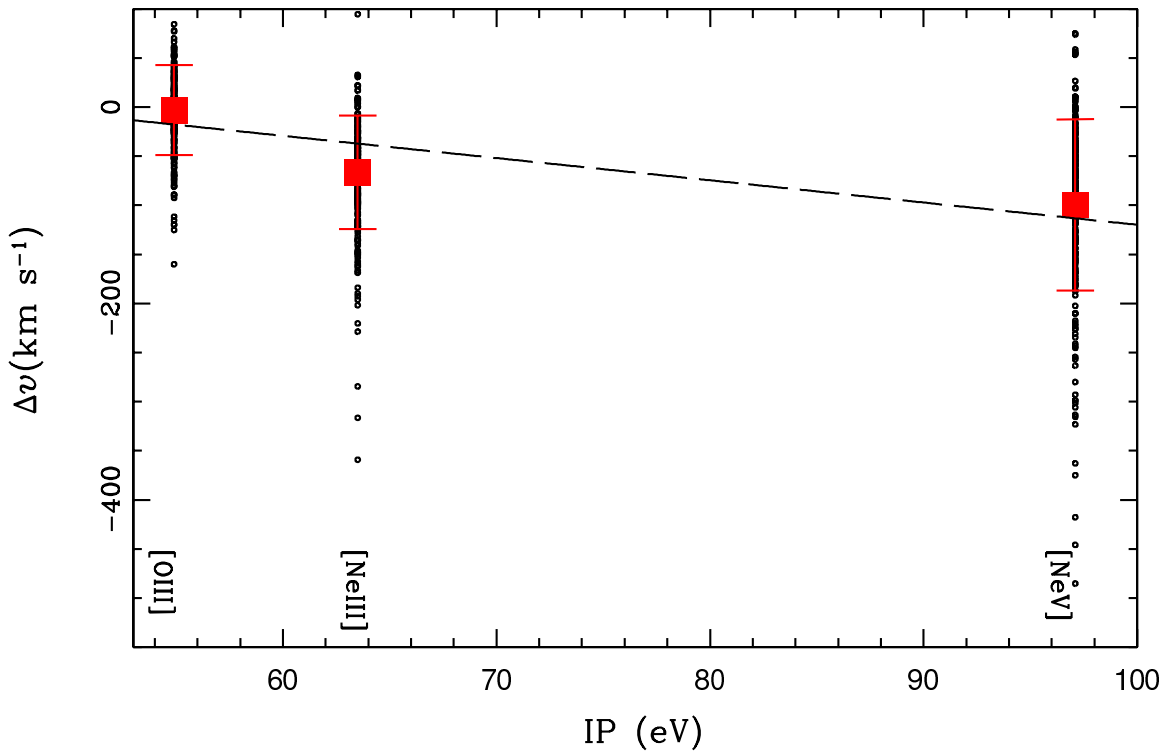}
\caption{An illustration of kinematic stratification of NLR gas by plotting the determined bulk velocity shift ($\Delta\upsilon$) against 
ionization potential (IP) for [\ion{O}{3}]$\lambda5007$, [\ion{Ne}{3}]$\lambda3869$ and [\ion{Ne}{5}]$\lambda3426$ emission lines. The small
dots represent the measurements of individual objects listed in our sample. The red squares and the corresponding uncertainties are the mean values 
and corresponding $1\sigma$ standard deviations. The best-fit line is overplotted by the dashed line.}
\end{figure}

\section{Discussion}

\subsection{Disk Wind/Radiation Origin of Outflow}

We identify a dependence of $L/L_{\mathrm{Edd}}$ on both blue wing of [\ion{O}{3}]$\lambda5007$ line and
bulk blueshift of high-ionized [\ion{Ne}{3}]$\lambda3869$ and [\ion{Ne}{5}]$\lambda3426$ for quasars at $z\sim0.4-0.8$,
which allows us to argue that $L/L_{\mathrm{Edd}}$ is the physical driver of the outflows occurring in the narrow-line region 
of AGNs up to $z\sim1$. 
The role of $L/L_{\mathrm{Edd}}$ on outflow has been, in fact, frequently revealed 
in nearby AGNs. 
Basing upon the large SDSS spectroscopic survey, a dependence of $L/L_{\mathrm{Edd}}$ on both 
[\ion{O}{3}] asymmetry and bulk blueshift has been identified in local luminous AGNs. 
Generally speaking, stronger the blue asymmetry and larger the bulk
velocity blueshift, higher the
$L/L_{\mathrm{Edd}}$ will be (e.g., Bian et al. 2005; Boroson 2005;
Zhang et al. 2011; Wang et al. 2011; Wang 2015).  
Matsuoka (2012) shows a deficit of the extended emission-line region in
the AGNs with high $L/L_{\mathrm{Edd}}$, which could be explained either by the AGN's outflow that
blows the gas around central SMBH away or by galaxy minor merger.
Wang et al. (2016) recently identified a
direct dependence of [\ion{O}{3}] blue asymmetry on the accretion process occurring around the central SMBH by
a joint X-ray and optical spectral analysis: stronger [\ion{O}{3}] blue asymmetry tends to occur in local AGNs with both 
steeper hard-X-ray spectra and higher $L/L_{\mathrm{Edd}}$.

In the theoretical ground, the outflow from luminous AGNs is 
proposed to be explained by the wind/radiation model. In the model,
the observed outflow is caused by the
wind/radiation pressure launched/emitted from the inner accretion disk 
(e.g., Murray et al. 1995; Proga et al. 2000;
Crenshaw et al. 2003; King \& Pounds 2003; Pounds
et al. 2003; King 2003, 2005; Ganguly et al. 2007; Reeves
et al. 2009; Alexander et al. 2010; Dunn et al. 2010; King
et al. 2011; Zubovas \& King 2012; Fabian 2012).
The disk wind model has been
successfully applied to explain the observed broad ultraviolet
absorption lines in a fraction of $\sim$20\% quasars and the ultrafast
outflows identified from the blueshifted X-ray \ion{Fe}{25} and
\ion{Fe}{26} absorptions in a few local AGNs (e.g., Tombesi
et al. 2012; Higginbottom et al. 2014). 
The hydrodynamic outflow model calculated by
Proga et al. (2008) indicates that the wind launched from the
accretion disk can extended into the inner NLR, though the
specific launch mechanism is still under debate. Possible
mechanisms include radiation/line-driven (e.g., Proga
et al. 1998, 2000; Laor \& Brandt 2002; Proga \& Kallman 2004;
Nomura et al. 2013; Higginbottom et al. 2014; Hagino et al.
2015), thermally driven (e.g., Begelman et al. 1983; Krolik \&
Kriss 2001), magnetically driven (e.g., Blandford \&
Payne 1982; Ferreira 1997; Fukumura et al. 2014; Stepanovs
\& Fendt 2014), and hybrid models (e.g., Proga 2003;
Everett 2005).

%\clearpage

%% The "ht!" tells LaTeX to put the figure "here" first, at the "top" next
%% and to override the normal way of calculating a float position

\subsection{Radio Emission\label{subsec:general}}

In addition to the wind/radiation model, another major scenario of outflow 
is that the outflow in NLR is driven by the interaction between the radio jet and
the interstellar medium (e.g., Heckman et al. 1984;
Whittle 1985, Brotherton 1996; Whittle \& Wilson 2004;
Morganti et al. 2007; Holt et al. 2008, 2011; Nesvadba
et al. 2008; Guillard et al. 2012; Mahony et al. 2013; Mullaney et al. 2013; 
Zakamska \& Greene 2014).

To examine the role of radio emission, we follow Wang et al. (2016) by 
cross-matching the $\sim900$ quasars at $z\sim 0.4-0.8$ with the FIRST survey
catalog (Becker et al. 2003). The cross-match with a radius of 3\symbol{125} returns 183 objects with detected 
radio emission exceeding the FIRST limiting flux density (5$\sigma$) of 1mJy.
The radio power at 1.4 GHz (rest frame) of each detection is obtained from the observed integrated flux density at 1.4 GHz
$f_\nu$ through $P_{\mathrm{1.4GHz}}=4\pi d_L^2 f_\nu(1+z)^{-1-\alpha}$,
where $d_L$ is the luminosity distance, and $\alpha=-0.8$ (e.g., Ker et al. 2012) is the
spectral slope defined as $f_\nu\propto\nu^{-\alpha}$.
For each of the objects without a detection, an upper limit of radio power 
is inferred from the reported detection limit at the corresponding celestial position through the same 
transformation.

The calculated $P_{\mathrm{1.4GHz}}$ is plotted as a function of [\ion{O}{3}] line asymmetry, bulk velocity offsets
of [\ion{O}{3}], [\ion{Ne}{3}] and [\ion{Ne}{5}] emission lines in Figure 6. 
In each panel, the blue points denote the
sources with a detected radio flux, and the red points the
ones with a flux upper limit. One can see from the figure that 
in the current sample both strength of the [\ion{O}{3}] blue wing and bulk velocity
blueshifts of [\ion{Ne}{3}] and [\ion{Ne}{5}] are unlikely driven by 
radio emission. This independence therefore allows us to exclude the 
radio jet origin of outflow in the current sample. In fact, the objects with strong
[\ion{O}{3}] blue asymmetry or large bulk velocity offsets instead tend to 
have either less or undetected radio power.

\begin{figure*}
\plotone{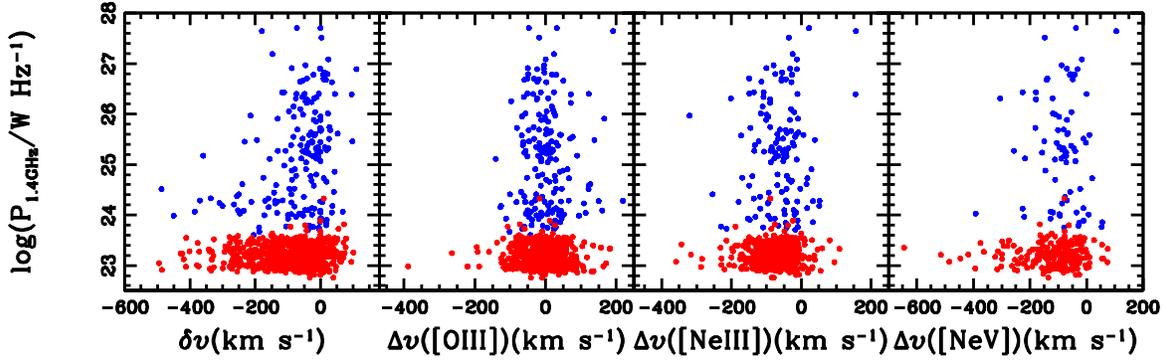}
\caption{The same as in Figure 2 but for 1.4GHz radio power at rest-frame. The sources
with a detected radio flux are shown by the blue points, and the ones with a flux upper limit by the red points.}
\end{figure*}

\subsection{Comparison to Local AGNs}

In this subsection, the quantified [\ion{O}{3}] line asymmetry based on the current sample is compared to
that of nearby X-ray selected type-I AGNs studied in Wang et al. (2016), simply because both studies adopt the same definition of line asymmetry quantification. 
In the following comparison, we ignore the possible systematics in $L/L_{\mathrm{Edd}}$ estimations caused by different calibrations:
the $L/L_{\mathrm{Edd}}$ is estimated by extinction-corrected H$\alpha$ broad emission in Wang et al. (2016) and 
by H$\beta$ in the current study. In fact, the effect on $L/L_{\mathrm{Edd}}$ due to extinction correction is quit small because 
$L/L_{\mathrm{Edd}}$ is scaled to line luminosity as $L/L_{\mathrm{Edd}}\sim L^{1/4}_{\mathrm{H\alpha,H\beta}}$.

The comparison between the two samples is illustrated in Figure 7. The lower two panels compare 
the cumulative distributions of $L_{\mathrm{H\beta}}$ and $M_{\mathrm{BH}}$. One can clearly see that, when compared to the sample of local AGNs, the 
quasar sample investigated in the current study is strongly biased towards the objects with high luminosity and large SMBH mass end. This significant 
discrepancy can be naturally explained by the detection limit effect in which only bright enough sources can be detected at higher redshift for 
a given flux limit.

The top-left and top-right panels compare the cumulative distributions of the quantified asymmetry and estimated $L/L_{\mathrm{Edd}}$ between the two samples,
respectively.
As shown by the comparisons, the local AGN sample is biased towards both stronger asymmetry and higher $L/L_{\mathrm{Edd}}$.
A two-side 1-dimensional Kolmogorov-Smirnov test shows a significant difference between the two samples. The mean values of $\delta\upsilon$ are
$-84\pm3$ and $-97\pm15\ \mathrm{km\ s^{-1}}$ for the intermediate-z quasars and nearby X-ray-selected AGNs, respectively.
The corresponding mean values of $L/L_{\mathrm{Edd}}$ are $0.25\pm0.03$ and $0.13\pm0.01$. This difference is not hard to be understood 
since the local AGNs in Wang et al. (2016) are X-ray selected, which biased against less active objects. The X-ray luminosities in
an energy band from 2 to 10 keV of these local X-ray selected AGNs range from $10^{42}$ to $10^{44}\ \mathrm{erg\ s^{-1}}$, 
and the $L/L_{\mathrm{Edd}}$ from 0.01 to 1. In fact, bright local AGNs usually have a $L/L_{\mathrm{Edd}}$ with a range of
at least 3 orders of magnitude from 0.001 to 1 (e.g., Woo \& Urry 2002; Boroson 2002). Basing upon the 2-10 keV X-ray
luminosity, the $L/L_{\mathrm{Edd}}$ of the Palomar optically selected AGNs ranges 
from $10^{-5}$ to 0.1 (Panessa et al. 2006) .

\begin{figure*}
\plotone{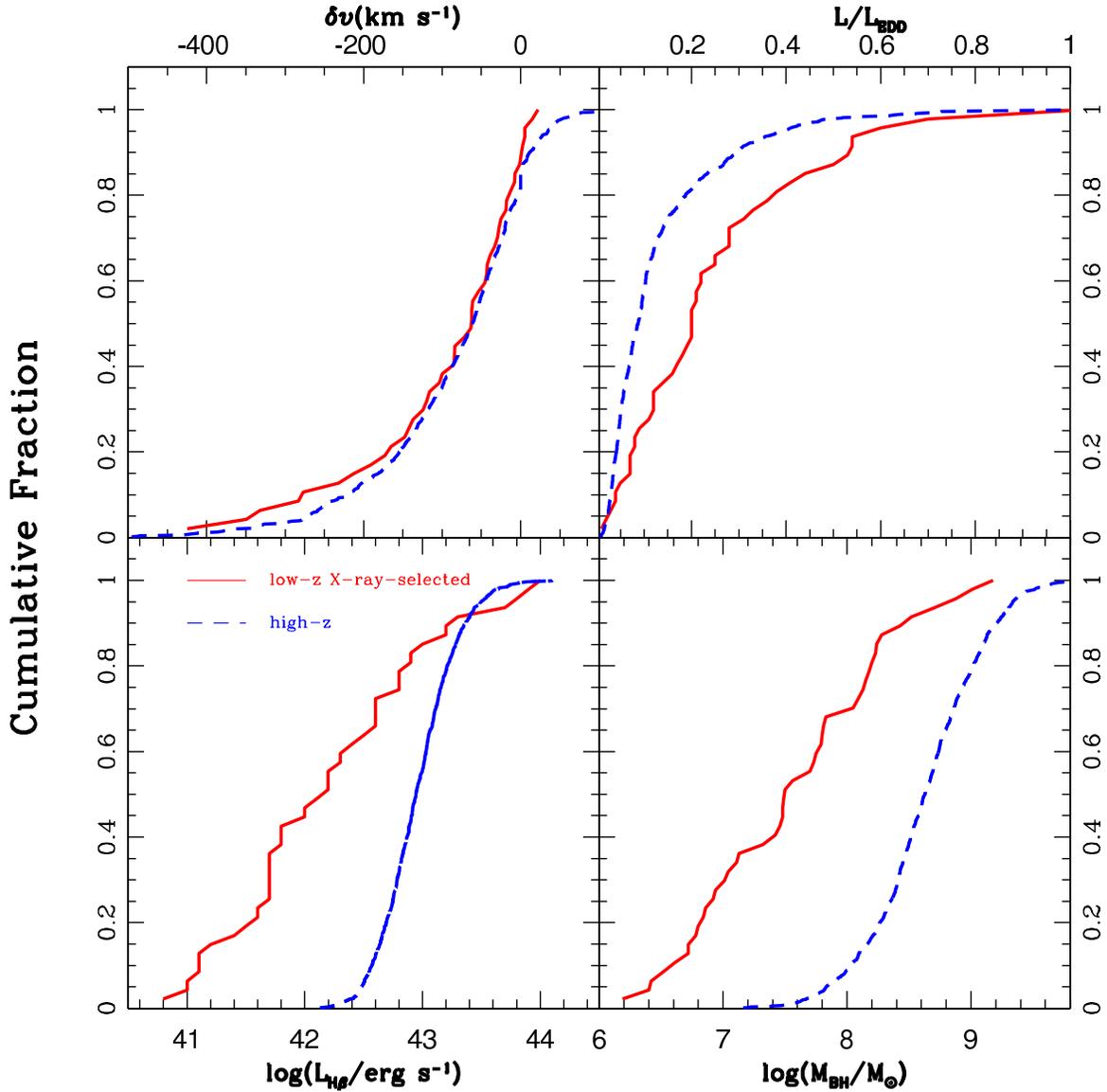}
\caption{Comparisons between the $z\sim0.4-0.8$ quasars of this study and the local X-ray-selected AGNs studied in Wang et al. (2016). 
The cumulative distributions of the $z\sim0.4-0.8$ quasars are shown by blue dashed lines, and those of 
local X-ray-selected AGNs by the red solid lines. }
\end{figure*}

Even though this difference, the intermediate-z quasars studied in this paper are combined with the local X-ray selected AGNs
studied in Wang et al. (2016) in Figure 8. The figure plots $L/L_{\mathrm{Edd}}$ against [\ion{O}{3}]$\lambda5007$ line asymmetry, which 
indicates that the local X-ray-selected AGNs are consistent with the $z\sim0.4-0.8$ quasars in terms of the $L/L_{\mathrm{Edd}}$
distribution as a function of the [\ion{O}{3}]$\lambda5007$ line asymmetry. 
A two-dimensional Kolmogorov-Smirnov test again yields a probability of $1.7\times10^{-5}$ that the two samples are 
drawn from the different parent populations.

% The dependence can be more clearly identified by both average and median values of $L/L_{\mathrm{Edd}}$ that are obtained 
% through a binning ($100\ \mathrm{km\ s^{-1}}$) of the combined sample with $\delta\upsilon$, which are overplotted in Figure 7 
% by the blue and green points and lines. 

\begin{figure}
\plotone{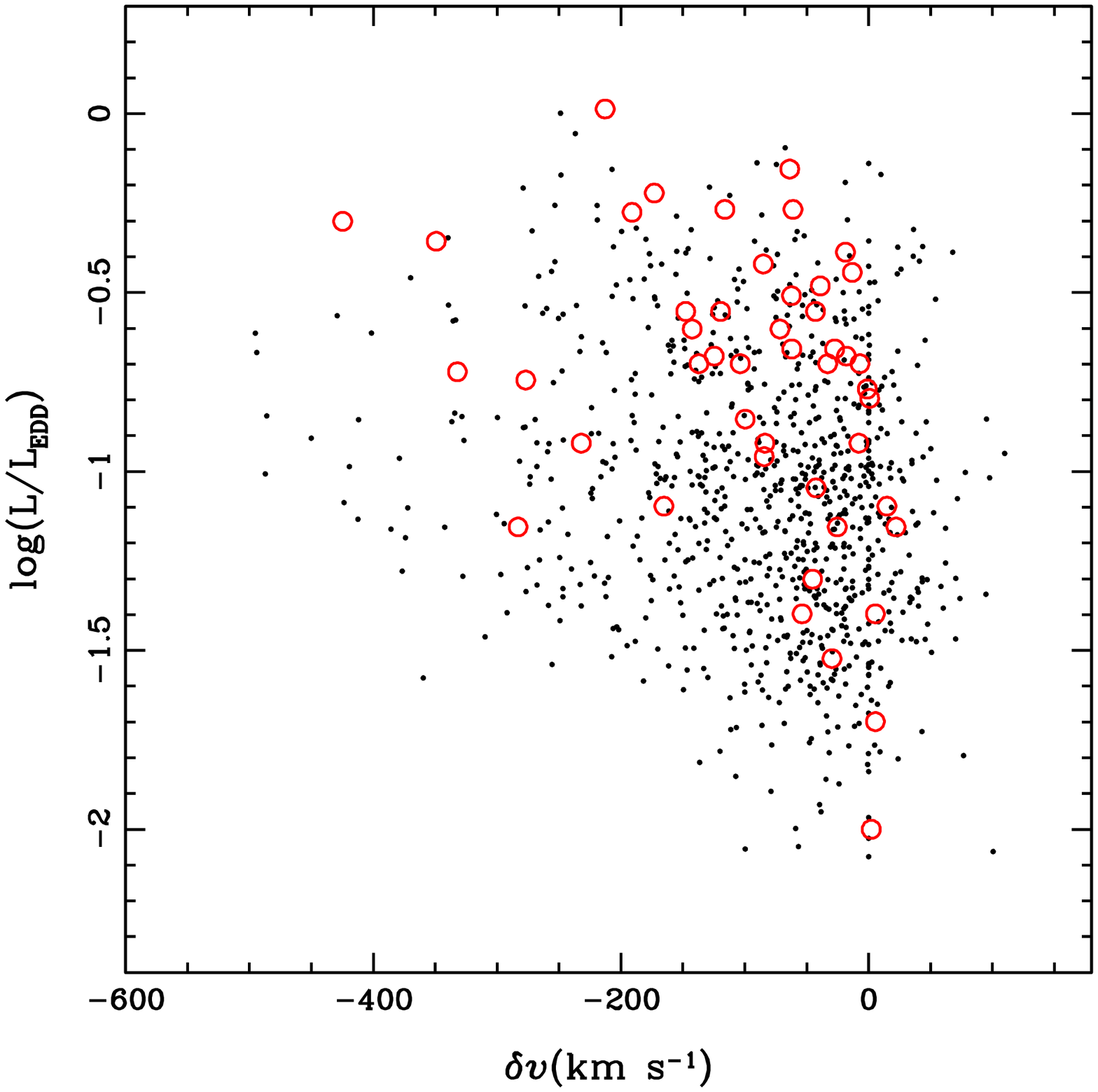}
\caption{[\ion{O}{3}]$\lambda5007$ emission line asymmetry $\delta\upsilon$ plotted against $L/L_{\mathrm{Edd}}$. The black points denote the 
measurements of the $z\sim0.4-0.8$ quasars obtained in this study, and the red open circles the measurements of the local X-ray-selected AGNs studied 
in Wang et al. (2016). 
%The blue and green symbols denote the average and median values (and corresponding deviations) of $L/L_{\mathrm{Edd}}$ 
%that are determined by a binning ($100\ \mathrm{km\ s^{-1}}$) with $\delta\upsilon$.
}
\end{figure}

\section{Conclusions} \label{sec:displaymath}

The outflow is systematically examined for a large sample of $\sim900$ quasars at $z\sim0.4-0.8$ selected from SDSS in this paper.
By modeling profiles of multiple emission lines by a sum of several Gaussian functions,
we identify a prevalence of not only [\ion{O}{3}]$\lambda5007$ line blue asymmetry, but also 
bulk velocity blueshift of both [\ion{Ne}{3}]$\lambda3869$ and [\ion{Ne}{3}]$\lambda3426$ lines, which are tend to be
associated with a high $L/L_{\mathrm{Edd}}$.  These results allow us to argue that  
the pressure caused by the wind/radiation launched/emitted from central SMBH is the most likely origin of the 
outflow in these distant quasars, which further implies that the outflow in luminous AGNs up to $z\sim1$ have the same origin.

%% Putting eqnarrays or equations inside the mathletters environment groups
%% the enclosed equations by letter. For instance, the eqnarray below, instead
%% of being numbered, say, (4) and (5), would be numbered (4a) and (4b).
%% LaTeX the paper and look at the output to see the results.

\acknowledgments
The authors would like to thank the anonymous referee for his/her very useful comments and suggestions for
improving the manuscript.
The study is supported by the National
Natural Science Foundation of China under grants 11473036
and 11773036, This study uses the SDSS archive data that was
created and distributed by the Alfred P. Sloan Foundation.

\end{document}